\documentclass[useAMS,usenatbib]{mnras}
\include{journals}
\usepackage{amsmath}
\usepackage{amssymb}
\usepackage{graphicx}

\newcommand\textlcsc[1]{\textsc{\MakeLowercase{#1}}}

\def\jref@jnl#1{{\rm#1}}

\def\aj{\jref@jnl{AJ}}                   
\def\araa{\jref@jnl{ARA\&A}}             
\def\apj{\jref@jnl{ApJ}}                 
\def\apjl{\jref@jnl{ApJ}}                
\def\apjs{\jref@jnl{ApJS}}               
\def\ao{\jref@jnl{Appl.~Opt.}}           
\def\apss{\jref@jnl{Ap\&SS}}             
\def\aap{\jref@jnl{A\&A}}                
\def\aapr{\jref@jnl{A\&A~Rev.}}          
\def\aaps{\jref@jnl{A\&AS}}              
\def\azh{\jref@jnl{AZh}}                 
\def\baas{\jref@jnl{BAAS}}               
\def\jrasc{\jref@jnl{JRASC}}             
\def\memras{\jref@jnl{MmRAS}}            
\def\mnras{\jref@jnl{MNRAS}}             
\def\pasa{\jref@jnl{PASA}}
\def\pra{\jref@jnl{Phys.~Rev.~A}}        
\def\prb{\jref@jnl{Phys.~Rev.~B}}        
\def\prc{\jref@jnl{Phys.~Rev.~C}}        
\def\prd{\jref@jnl{Phys.~Rev.~D}}        
\def\pre{\jref@jnl{Phys.~Rev.~E}}        
\def\prl{\jref@jnl{Phys.~Rev.~Lett.}}    
\def\pasp{\jref@jnl{PASP}}               
\def\pasj{\jref@jnl{PASJ}}               
\def\qjras{\jref@jnl{QJRAS}}             
\def\skytel{\jref@jnl{S\&T}}             
\def\solphys{\jref@jnl{Sol.~Phys.}}      
\def\sovast{\jref@jnl{Soviet~Ast.}}      
\def\ssr{\jref@jnl{Space~Sci.~Rev.}}     
\def\zap{\jref@jnl{ZAp}}                 
\def\nat{\jref@jnl{Nature}}              
\def\iaucirc{\jref@jnl{IAU~Circ.}}       
\def\aplett{\jref@jnl{Astrophys.~Lett.}} 
\def\apspr{\jref@jnl{Astrophys.~Space~Phys.~Res.}}
\def\bain{\jref@jnl{Bull.~Astron.~Inst.~Netherlands}} 
\def\fcp{\jref@jnl{Fund.~Cosmic~Phys.}}  
\def\gca{\jref@jnl{Geochim.~Cosmochim.~Acta}}   
\def\grl{\jref@jnl{Geophys.~Res.~Lett.}} 
\def\jcp{\jref@jnl{J.~Chem.~Phys.}}      
\def\jgr{\jref@jnl{J.~Geophys.~Res.}}    
\def\jqsrt{\jref@jnl{J.~Quant.~Spec.~Radiat.~Transf.}}
\def\memsai{\jref@jnl{Mem.~Soc.~Astron.~Italiana}}
\def\nphysa{\jref@jnl{Nucl.~Phys.~A}}   
\def\physrep{\jref@jnl{Phys.~Rep.}}   
\def\physscr{\jref@jnl{Phys.~Scr}}   
\def\planss{\jref@jnl{Planet.~Space~Sci.}}   
\def\procspie{\jref@jnl{Proc.~SPIE}}   

\usepackage{datatool}

\title[Chemically distinct discs]{Origin of chemically distinct discs in the Auriga cosmological simulations}
\author[Grand et al.]{\parbox[t]{\textwidth}{
Robert J. J. Grand$^{1,2}$\thanks{robert.grand@h-its.org}, Sebasti\'{a}n Bustamante$^1$, Facundo A. G\'{o}mez$^{3,4,5}$, Daisuke Kawata$^6$, Federico Marinacci$^7$, R\"{u}diger Pakmor$^1$, Hans-Walter Rix$^8$, Christine M. Simpson$^1$, Martin Sparre$^9$, Volker Springel$^{1,2}$ } \vspace{10pt} \\
$^1$Heidelberger Institut f\"{u}r Theoretische Studien, Schloss-Wolfsbrunnenweg 35, 69118 Heidelberg, Germany\\
$^2$Zentrum f\"{u}r Astronomie der Universit\"{a}t Heidelberg, Astronomisches Recheninstitut, M\"{o}nchhofstr. 12-14, 69120 Heidelberg, Germany\\
$^{3}$Instituto de Investigaci{\'o}n Multidisciplinar en Ciencia yTecnolog{\'i}a, Universidad de La Serena, Ra{\'u}l Bitr{\'a}n 1305, La Serena, Chile\\
$^{4}$Departamento de F{\'i}sica y Astronom{\'i}a, Universidad de LaSerena, Av. Juan Cisternas 1200 N, La Serena, Chile\\
$^5$Max-Planck-Institut f\"{u}r Astrophysik, Karl-Schwarzschild-Str. 1, D-85748, Garching, Germany  \\
$^6$Mullard Space Science Laboratory, University College London, Holmbury St. Mary, Dorking, Surrey, RH5 6NT, United Kingdom \\
$^7$Department of Physics, Kavli Institute for Astrophysics and Space Research, MIT, Cambridge, MA 02139, USA \\
$^8$ Max Planck Institute for Astronomy, K\"{o}nigstuhl 17, D-69117 Heidelberg, Germany\\
$^9$Leibniz-Institute f\"{u}r Astrophysik Potsdam (AIP), An der Sternwarte16, D-14482 Potsdam, Germany\\
}

\date{Accepted XXX. Received YYY; in original form ZZZ}

\pubyear{2017}
    
\begin{document}

\label{firstpage}

\pagerange{\pageref{firstpage}--\pageref{lastpage}}
\maketitle

\begin{abstract}
The stellar disk of the Milky Way shows complex spatial and abundance structure that is central to understanding the key physical mechanisms responsible for shaping our Galaxy. In this study, we use six very high resolution cosmological zoom simulations of Milky Way-sized haloes to study the prevalence and formation of chemically distinct disc components. We find that our simulations develop a clearly bimodal distribution in the $[\rm \alpha/Fe]$ -- $[\rm Fe/H]$ plane. We find two main pathways to creating this dichotomy which operate in different regions of the galaxies: a) an early ($z>1$) and intense high-$\rm[\alpha/Fe]$ star formation phase in the inner region ($R\lesssim 5$ kpc) induced by gas-rich mergers, followed by more quiescent low-$\rm[\alpha/Fe]$ star formation; and b) an early phase of high-$\rm[\alpha/Fe]$ star formation in the outer disc followed by a shrinking of the gas disc owing to a temporarily lowered gas accretion rate, after which disc growth resumes. In process b), a double-peaked star formation history around the time and radius of disc shrinking accentuates the dichotomy. If the early star formation phase is prolonged (rather than short and intense), chemical evolution proceeds as per process a) in the inner region, but the dichotomy is less clear. In the outer region, the dichotomy is only evident if the first intense phase of star formation covers a large enough radial range before disc shrinking occurs; otherwise, the outer disc consists of only low-$\rm[\alpha/Fe]$ sequence stars. We discuss the implication that both processes occurred in the Milky Way. 
\end{abstract}

\begin{keywords}
galaxies: evolution - galaxies: kinematics and dynamics - galaxies: spiral - galaxies: structure
\end{keywords}

\section{Introduction}

For many years, our understanding of the formation history of the Galaxy has heavily drawn on the concept of distinct thin and thick stellar disc components \citep[see][for a summary of the current status]{C17}. A structural dichotomy was first surmised from star counts in the solar neighbourhood, which were found to make up a vertical density distribution consistent with a sum of two exponential profiles of different scale heights \citep{GR83}. The idea of two distinct components was then bolstered by the distribution of solar neighbourhood disk stars in the $\rm [\alpha/Fe]$ -- $\rm [Fe/H$] plane \citep{GCM96,F98}: a distinct group of high- and low-$\rm[\alpha/Fe]$ stars with a dearth of stars in between. Recent Galactic spectroscopic surveys such as the Apache Point Observatory Galactic Evolution Experiment (APOGEE) have extended our view of the Galaxy and have revealed that the high-$\rm[\alpha/Fe]$ component is thicker and more radially compact than the low-$\rm[\alpha/Fe]$ component \citep{RTL03,BAO11,BP12,ACS14,HBH15}. The radial truncation of the high-$\rm[\alpha/Fe]$ component at approximately the solar radius indicates that there is no simple mapping from the bimodal structure in abundance space to real space; the structural thick disc outside of the solar radius cannot be part of the high-$\rm[\alpha/Fe]$ component - the chemically defined thick disc. Instead, the current picture is that the low-$\rm[\alpha/Fe]$ component is composed of many sub-populations of unique metallicity \citep[][]{BRS15,MSC17,MBS17} - the so-called Mono-Abundance Populations (MAPs), each of which are distributed in a ``donut'' morphology that becomes radially larger and  vertically thicker with decreasing metallicity. Moreover, the vertical scale height of each MAP flares (scale height increases with radius), with the flare becoming more emphatic with decreasing metallicity \citep{BRS15}. Similar flaring patterns have also been reported for coeval stellar populations in simulations \citep{MSC17}. In contrast, the high-$\rm[\alpha/Fe]$ MAPs are inferred to have a centrally peaked surface density with similar exponential scale radii and non-flaring scale heights, suggestive of a different origin to the low-$\rm[\alpha/Fe]$ component \citep[e.g.][]{HDS15}.
 
There are several proposed scenarios for creating a structurally distinct thick disc, including vertical heating from nearby satellites \citep{QHF93,VH08}, turbulence in high-redshift gas-rich discs \citep{N98,BEM09}, violent gas-rich mergers \citep{BKG04} and the accretion of satellite stars \citep{ANS03b}. More recently, it has been proposed that a distinct thick disc can be generated by the radial migration of kinematically hot stars from the inner to the outer disc \citep{SB09b,LRD11,Rok12}. While radial migration is expected to be important to reproduce the flat age-metallicity relation \citep{CSA11} and the metallicity-rotation velocity relation \citep{AKC16,KWC17,SM17}, it is still hotly debated whether it plays a key role in generating a thick disc \citep{Min12,KPA13,VC14,GSG16,KGG17}.

A promising explanation for both the geometrical and chemical properties of these two components is characterised by two phases of star formation: an early phase from gas infall during a rapid star formation phase to form a thick disc followed by gradual infall of primordial gas to form a thin disc. \citet{CMG97} first devised a two-infall analytical chemical evolution framework to model this scenario, which provided a good fit to the dichotomy in the local metallicity distribution, under the assumption that metal-rich, high-$\rm[\alpha/Fe]$ stars migrated from the inner to the outer disc \citep[see][for an alternative explanation based on parallel evolution]{GSM17}. However, these models could not account for the extremely metal-rich stars with thin disk kinematics at the local volume, which were shown to have a broad range in age (\citealt{TBE11,CSA11} and discussion in Section 4 of \citealt{ACR17}).

In recent years, numerical simulations have improved sufficiently to study the formation of thin and thick discs \citep[e.g.,][to name but a few]{RKA11,SWS11,LRD11,OSD16,MPG16,BMO17,MHW17,NYL17}. In particular, \citet{BKG04} employed cosmological zoom simulations to show that infalling gas brought into the galaxy by violent gas rich mergers can create a compact, centralised component \citep[see also][]{BRK07,RBM10} distinct from the comparably quiescent subsequent formation of a younger, lower-$\rm[\alpha/Fe]$ component. More modern cosmological simulations that better resolve the disc seem to reinforce this picture, and consistently predict that the structural thin disc forms smoothly in an inside-out, upside-down fashion \citep[e.g.][]{SBR13,BK13,MCM14,GSG16,DOB17}. The chemical dichotomy, however, does not seem to be a ubiquitous feature of simulations; \citet{BSG12} finds some semblance of two distinct sequences in chemical abundance space, whereas the evolution is smooth and continuous in the simulations of \citet{MCM13}. Thus, the formation of a clear chemical dichotomy is not guaranteed in simulations, which raises interesting questions regarding the formation mechanisms of the observed chemical dichotomy and whether it can be commonly reproduced in simulations.

In this paper, we analyse the Auriga simulation suite - a large set of high resolution, cosmological simulations of Milky Way-mass haloes \citep[see][]{GGM17}, which provide sufficient resolution to study the structure of the disc(s) in chemical abundance space in a fully cosmological context. A brief description of our simulations is given in Sec.~\ref{sec2}. In particular, we focus on the evolution of the $\rm [\alpha / Fe]-[Fe/H]$ plane of star particles in the disc, as a function of radius and height above the plane. Our goal is to search for distinct chemical components. We identify the prevalence of such components and determine the factors relevant for their development in Sec.~\ref{sec3}. We summarise out results and discuss the implications for the Milky Way formation history in Sec.~\ref{sec4}.

\section{Simulations}
\label{sec2}

The Auriga simulations \citep{GGM17} are a suite of cosmological zoom simulations of haloes in the mass range $10^{12}$ - $2\times10^{12}$ $\rm M_{\odot}$. The simulations begin at $z=127$ with cosmological parameters:  $\Omega _m = 0.307$, $\Omega _b = 0.048$, $\Omega _{\Lambda} = 0.693$ and a Hubble constant of $H_0 = 100 h$ km s$^{-1}$ Mpc$^{-1}$, where $h = 0.6777$, taken from \citet{PC13}. The simulations are performed with the magneto-hydrodynamic code \textlcsc{AREPO} \citep{Sp10}, with a comprehensive galaxy formation model \citep[see][for more details]{VGS13,MPS14,GSG16} that includes: primordial and metal line cooling, a prescription for a uniform background UV field for reionization (completed at $z=6$), a subgrid model for star formation, magnetic fields, and black hole seeding, accretion and feedback. For stellar feedback, a gas cell eligible for star formation is chosen (according to the \citet{C03} Initial Mass Function) to be either a site for SNII feedback or converted into a star particle. In the former case, a single wind particle is launched with a velocity that scales with the 1D local dark matter velocity dispersion in a random direction. Its metal content is determined by the initial metallicity of the gas cell that spawned the wind particle, the newly produced metals derived from the yield tables of \citet{PCB98}, and the metal loading factor, $\eta =0.6$: the wind particle is loaded with $(1-\eta)$ of the total metals, and the rest remains local. In the latter case, a star particle is created, which represents a single stellar population with a given mass, age and metallicity. Mass loss and metal enrichment from SNIa and AGB stars are modelled by calculating at each time step the mass moving off the main sequence for each star particle \citep[see][for more details]{GGM17}. The mass and metals are then distributed among nearby gas cells with a top-hat kernel. We track a total of 9 elements: H, He, C, O, N, Ne, Mg, Si and Fe.

In this study, we focus on the 6 highest resolution simulations in the suite. The typical dark matter particle mass is $\sim 4 \times 10^{4}$ $\rm M_{\odot}$, and the baryonic mass resolution is $\sim 5 \times 10^{3}$ $\rm M_{\odot}$. The physical softening of collisionless particles grows with time (corresponding to a fixed comoving softening length of 250 pc $h^{-1}$) until a maximum physical softening length of 185 pc is reached. The physical softening value for the gas cells is scaled by the gas cell radius (assuming a spherical cell shape given the volume), with a minimum softening set to that of the collisionless particles. We reiterate that in this study we analyse the 2-dimensional distribution of stars in chemical space in kiloparsec-scale spatial regions of the disc, which requires of order $10^7$ star particles in the main galaxy in order to sufficiently resolve the features for which we are searching.


\section{Results}
\label{sec3}

\begin{table}
\centering
\caption{Table of properties of the most significant mergers for each simulation. The columns are: 1) halo number; 2) subhalo number; 3) the merger time; 4) stellar mass ratio between the satellite and main galaxy at the time the satellite reaches its maximum stellar mass; 5) gas mass ratio between the satellite and main galaxy at the same time; 6) the gas mean metallicity ratio between the satellite and main galaxy at the same time.}
\label{t1}
\begin{tabular}{c c c c c c}
\hline
Run & Subhalo $\#$ & $\rm t_{m}$ [Gyr] & $\rm M_{*,rat}$ & $\rm M_{g,rat}$ & $\rm Z_{g,rat}$  \\
\hline                                                                     
Au 6    & 1 & 10.84 & 0.54 & 0.18 & 0.93 \\
Au 16 & 1 & 9.60    & 0.09 & 0.12 & 0.32 \\
Au 21 & 1 & 7.46    & 0.53 & 0.23 & 1.42 \\
	   & 2 & 7.74   & 0.40 & 0.25 & 0.59 \\
Au 23  & 1 & 11.74 & 0.34 & 0.21 & 0.87 \\
	   & 2 & 9.92   & 0.34 & 0.45 & 0.59 \\
Au 24  & 1 & 10.23 & 0.40 & 0.20 & 1.03 \\
Au 27  & 1 & 10.23 & 0.27 & 0.23 & 0.57 \\
	   & 2 & 9.31   & 0.27 & 0.21 & 0.74 \\

\hline
\end{tabular}
\end{table}

\begin{figure*}
\includegraphics[scale=0.9,trim={1.4cm 0.8cm .5cm 0},clip]{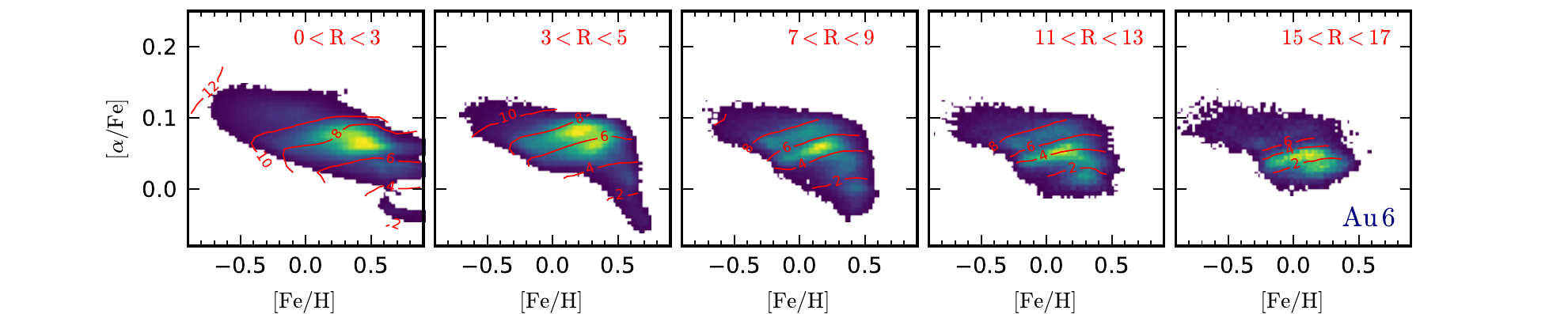}
\includegraphics[scale=0.9,trim={1.4cm 0.8cm .5cm 0},clip]{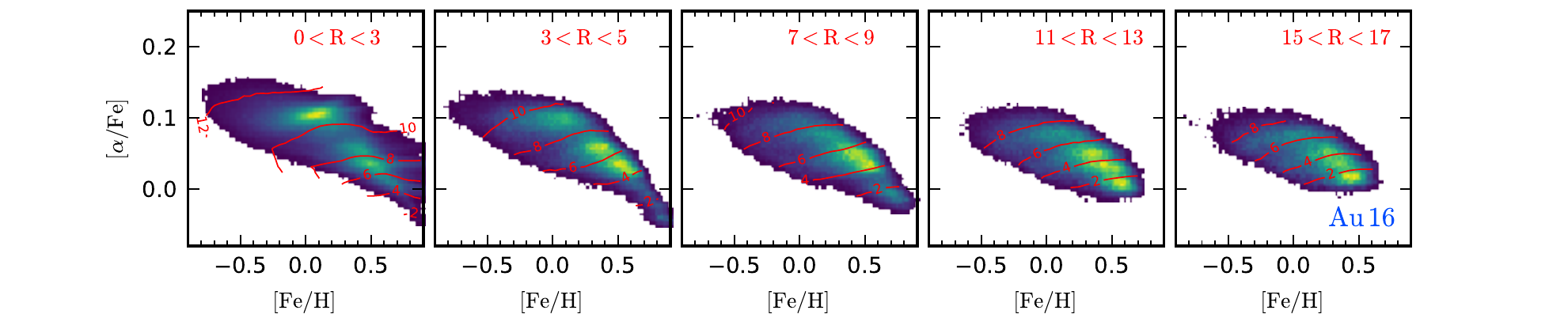}\\
\includegraphics[scale=0.9,trim={1.4cm 0.8cm .5cm 0},clip]{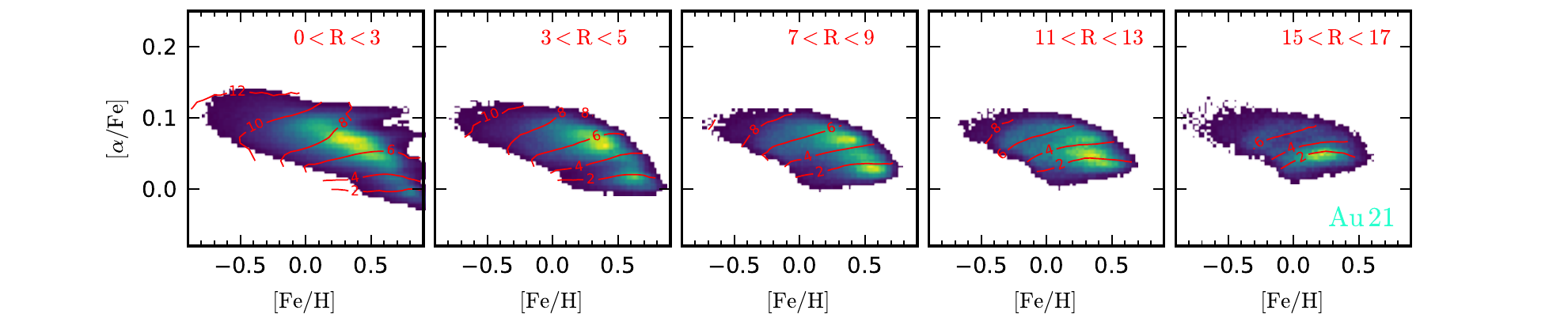}
\includegraphics[scale=0.9,trim={1.4cm 0.8cm .5cm 0},clip]{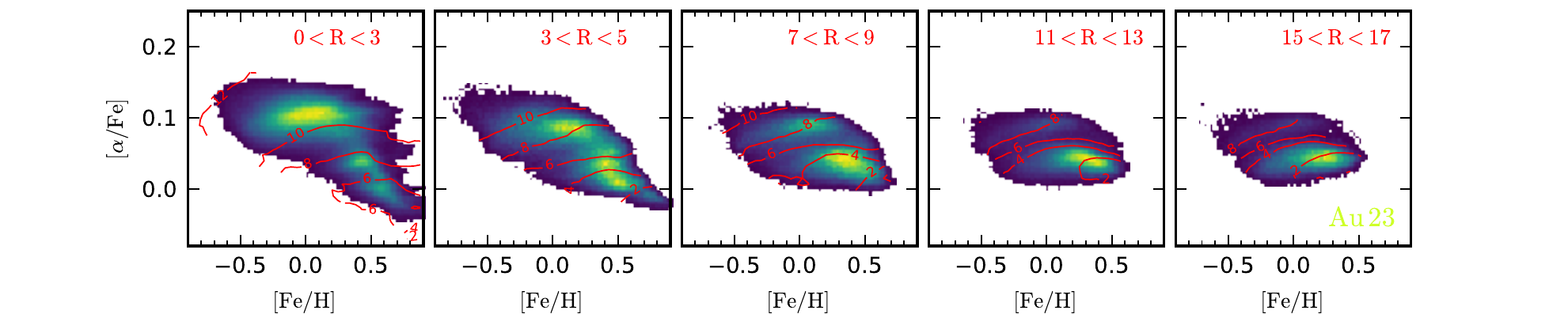}\\
\includegraphics[scale=0.9,trim={1.4cm 0.8cm .5cm 0},clip]{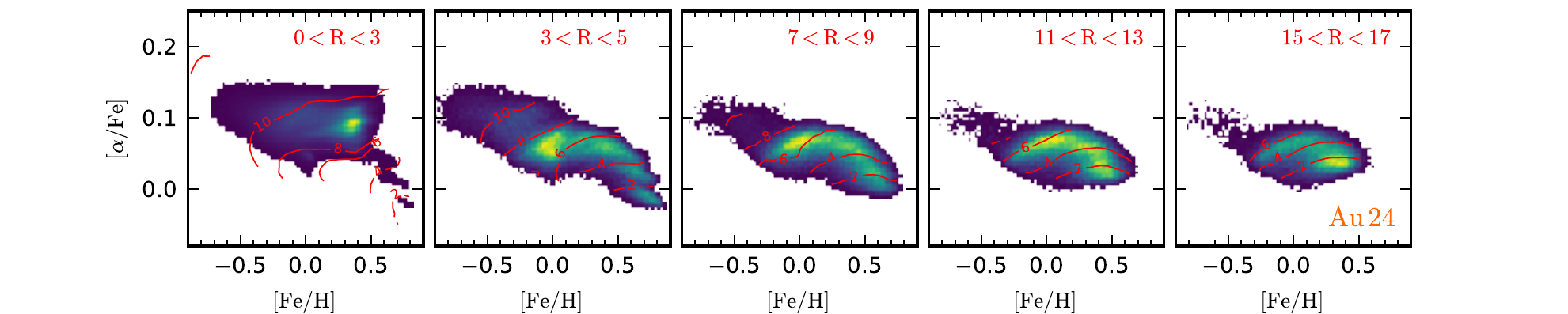}
\includegraphics[scale=0.9,trim={1.4cm 0.cm .5cm 0},clip]{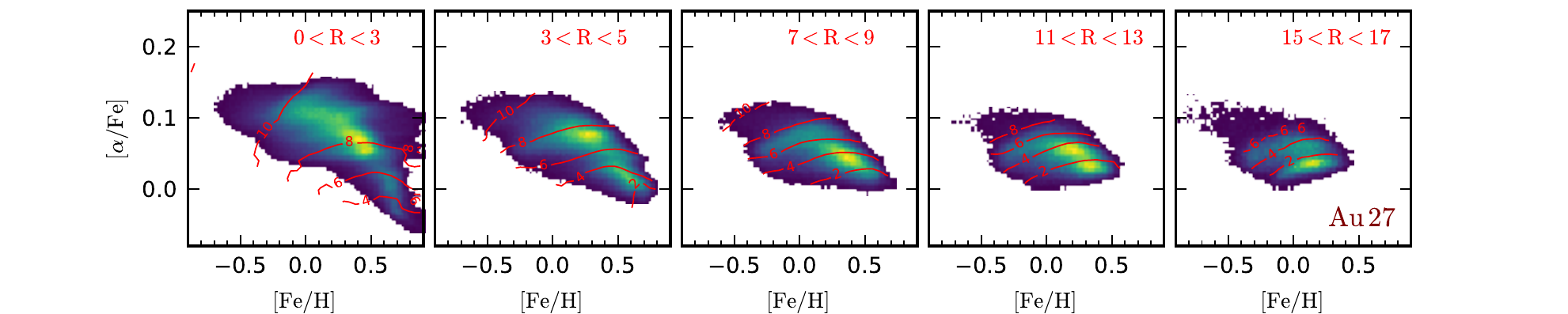}\\
\caption{The normalised number density of all star particles in the $[\rm \alpha/Fe]$ -- $[\rm Fe/H]$ plane, at different galactocentric radii, $R$, in the disc plane, for all simulations. Contours mark the regions of mean stellar age (all pixels above a contour have a mean age greater than that indicated on the contour):  evolution proceeds toward lower $[\rm \alpha/Fe]$ and higher $[\rm Fe/H]$ values. Distinct high-$\rm[\alpha/Fe]$ and low-$\rm[\alpha/Fe]$ components are visible at some radii in some simulations, whereas in others the distribution is smooth. The histograms are linear in scale and normalised to one in each panel in order to elucidate the features at each radius.}
\label{fig1}
\end{figure*}

\begin{figure*}
\includegraphics[scale=0.81,trim={4.cm 0.4cm 0 0},clip]{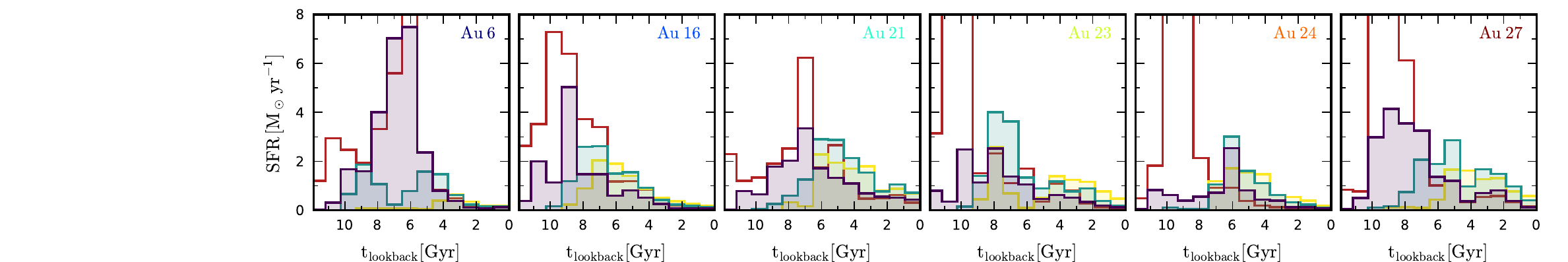}
\includegraphics[scale=0.81,trim={3.15cm 0.8cm -1.5cm 0.},clip]{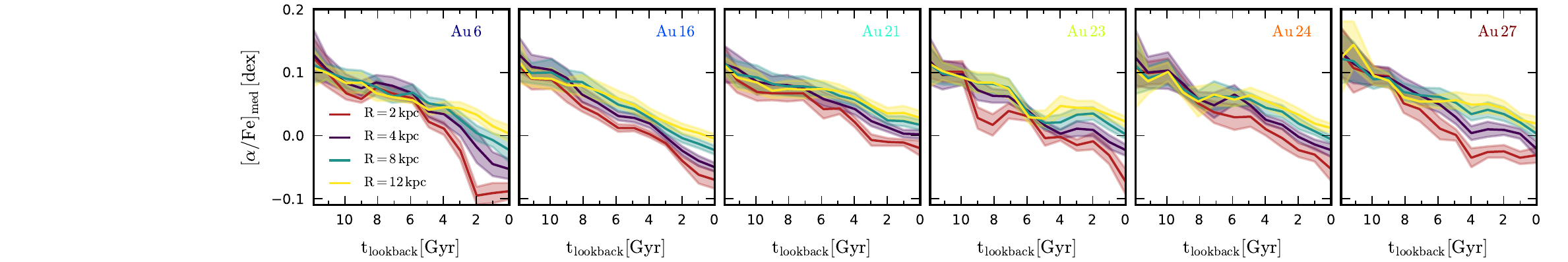}
\includegraphics[scale=1.055,trim={3.55cm 0 0.4cm 0.5cm},clip]{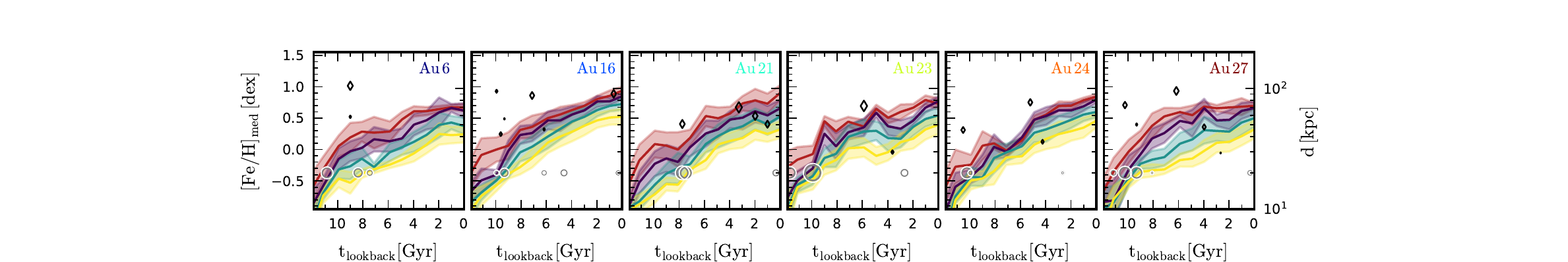}
\caption{\emph{Top-panels}: The star formation histories (SFHs) of all simulations in the radial regions $R<3$ kpc (red), and three 1 kpc-wide radial annuli centred on the radii $R=4$ kpc, 8 kpc and 12 kpc, and within 2 kpc of the midplane. Note that the peaks in the inner regions in some cases extend beyond the $y$-axis limit. \emph{Middle-panels}: The evolution of the median gas $\rm [\alpha/Fe]$ distribution (solid curve) in the same radial annuli as the top panels. The shaded region indicates the root mean square of the gas $\rm [\alpha/Fe]$ distribution. \emph{Bottom-panels}: As the middle panels, but for the gas $\rm[Fe/H]$. The distances of flybys (diamonds) are indicated on the right-hand axis, and circles indicate the times of gas-rich mergers (listed in Table.~1), plotted at $\rm d=20$ kpc for visual convenience. The size of the circles is proportional to the gas mass ratio of the satellites to the main galaxy, defined at the time when the satellite reaches its peak stellar mass, whereas the size of the diamonds is proportional to the instantaneous gas mass ratio for that time.}
\label{fig2}
\end{figure*}

\begin{figure}
\includegraphics[scale=1.8,trim={0 0 0 0},clip]{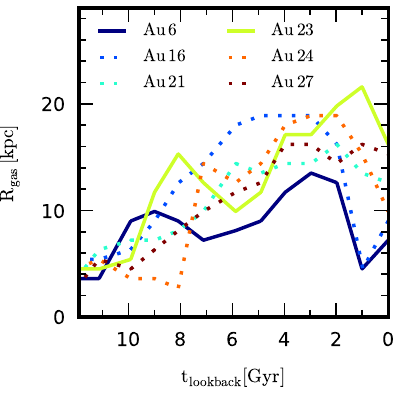}
\caption{The evolution of the `edge' of the gas disc, defined to be the radius at which the gas surface density falls below $5$ $\rm M_{\odot}$ $\rm pc^{-2}$, for all simulations. Each curve is coloured according to the label in Fig.~\ref{fig2}.}
\label{sfg}
\end{figure}

\begin{figure*}
\includegraphics[scale=0.6,trim={1.cm 1.cm 2.5cm 0},clip]{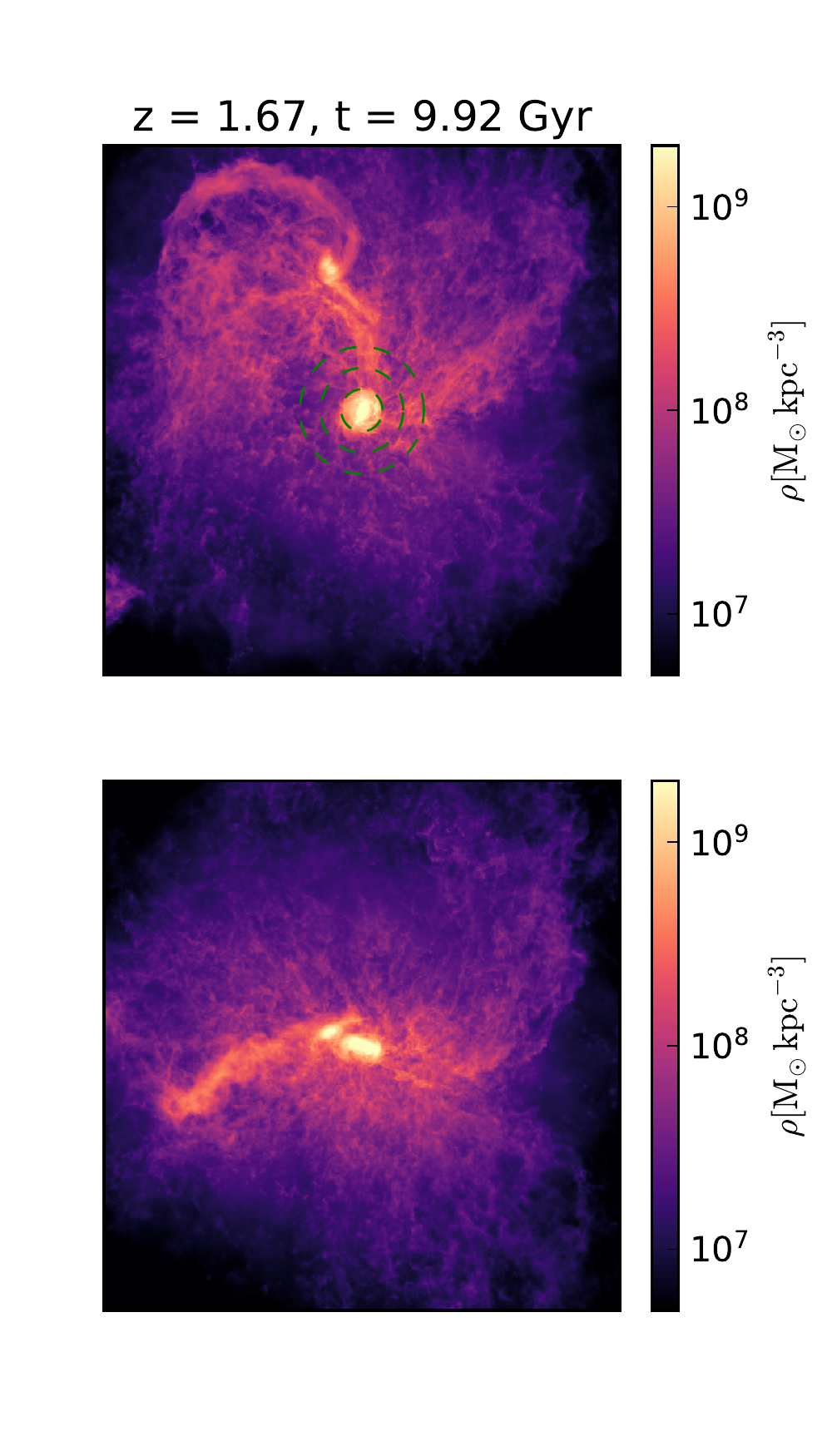}
\includegraphics[scale=0.6,trim={1.cm 1.cm 2.5cm 0.cm},clip]{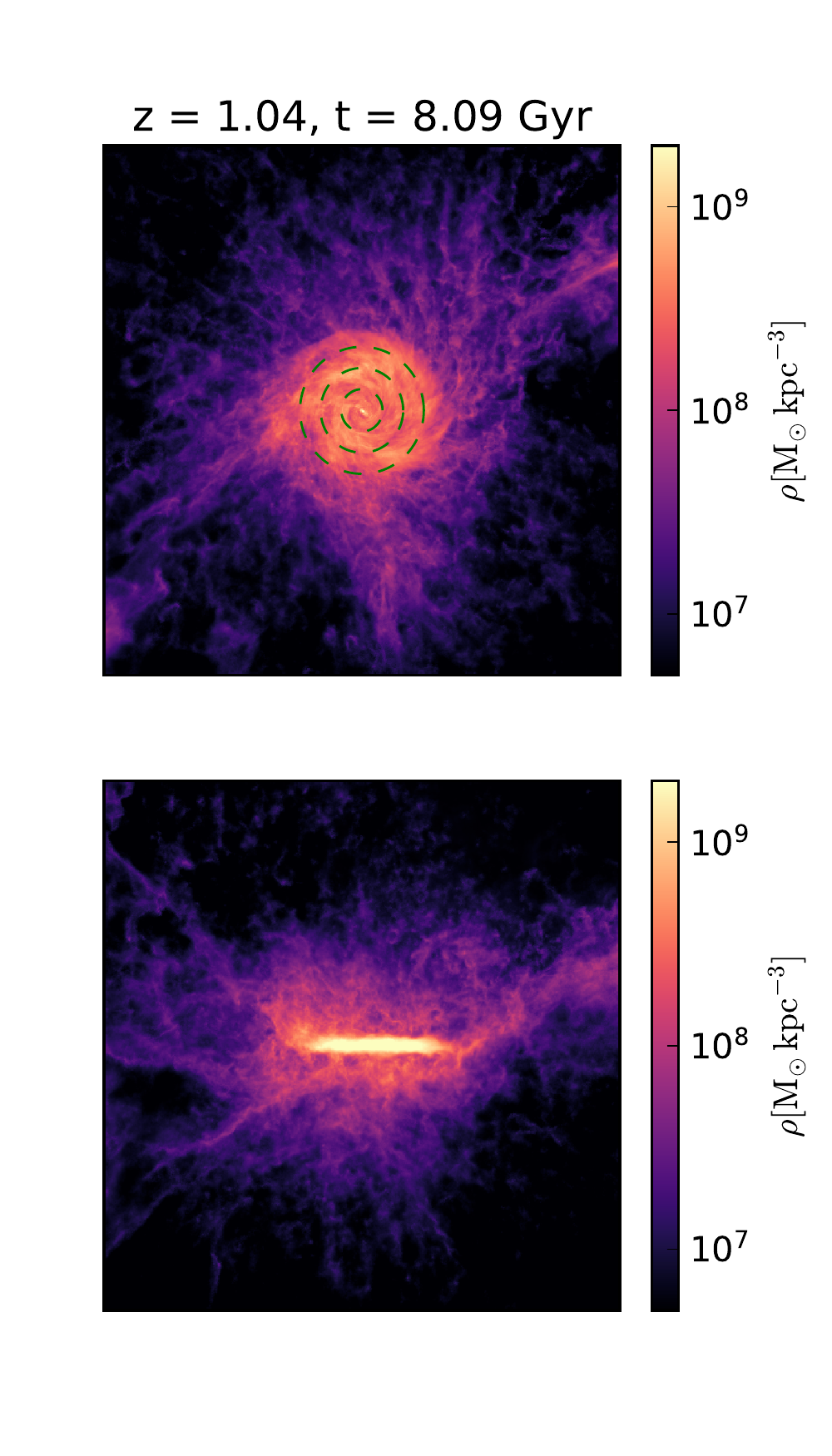}
\includegraphics[scale=0.6,trim={1.cm 1.cm 2.5cm 0.cm},clip]{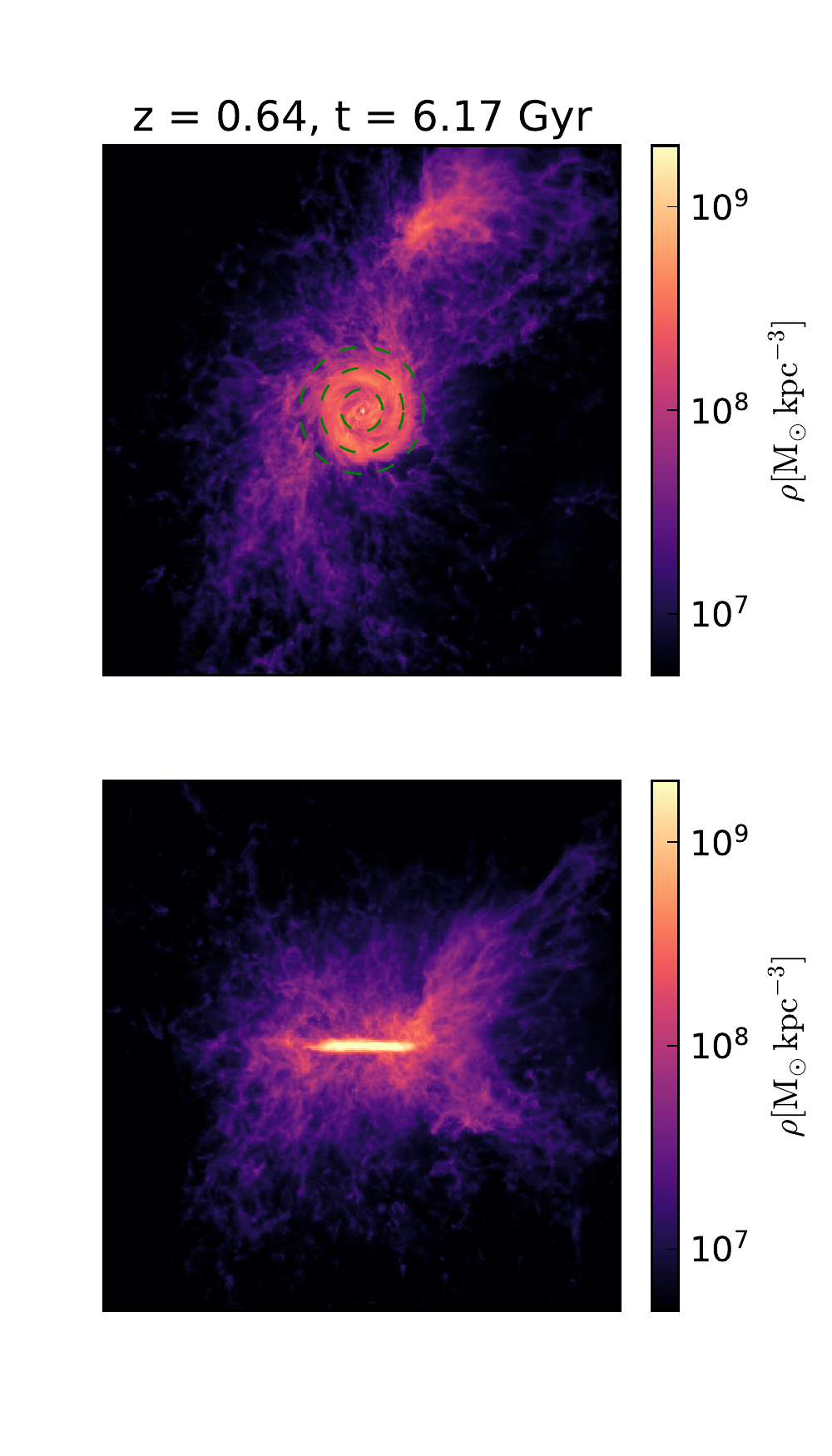}
\includegraphics[scale=0.6,trim={1.cm 1.cm 0.2cm 0.cm},clip]{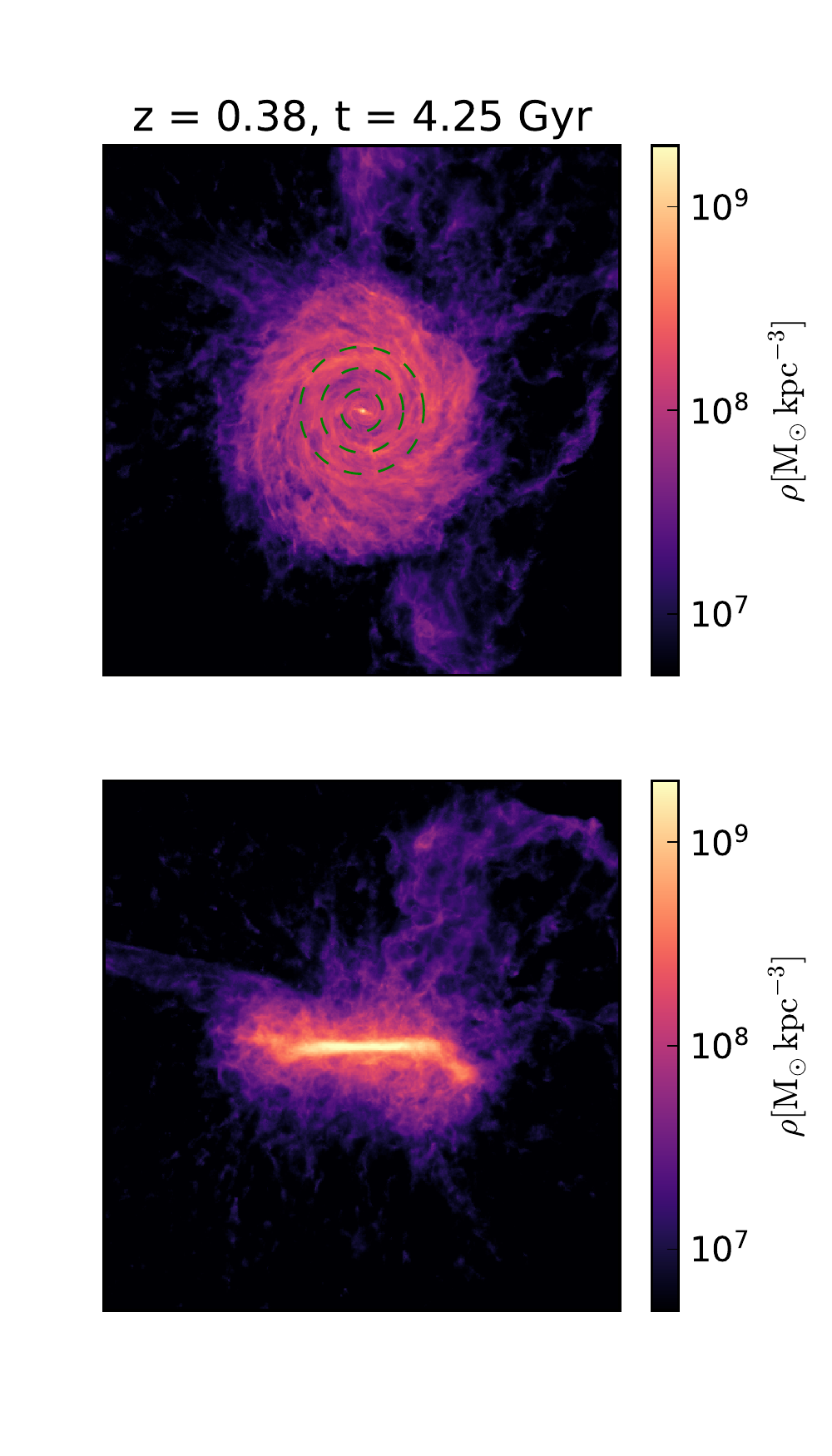}\\
\includegraphics[scale=0.6,trim={1.cm 1.cm 2.5cm 1.75cm},clip]{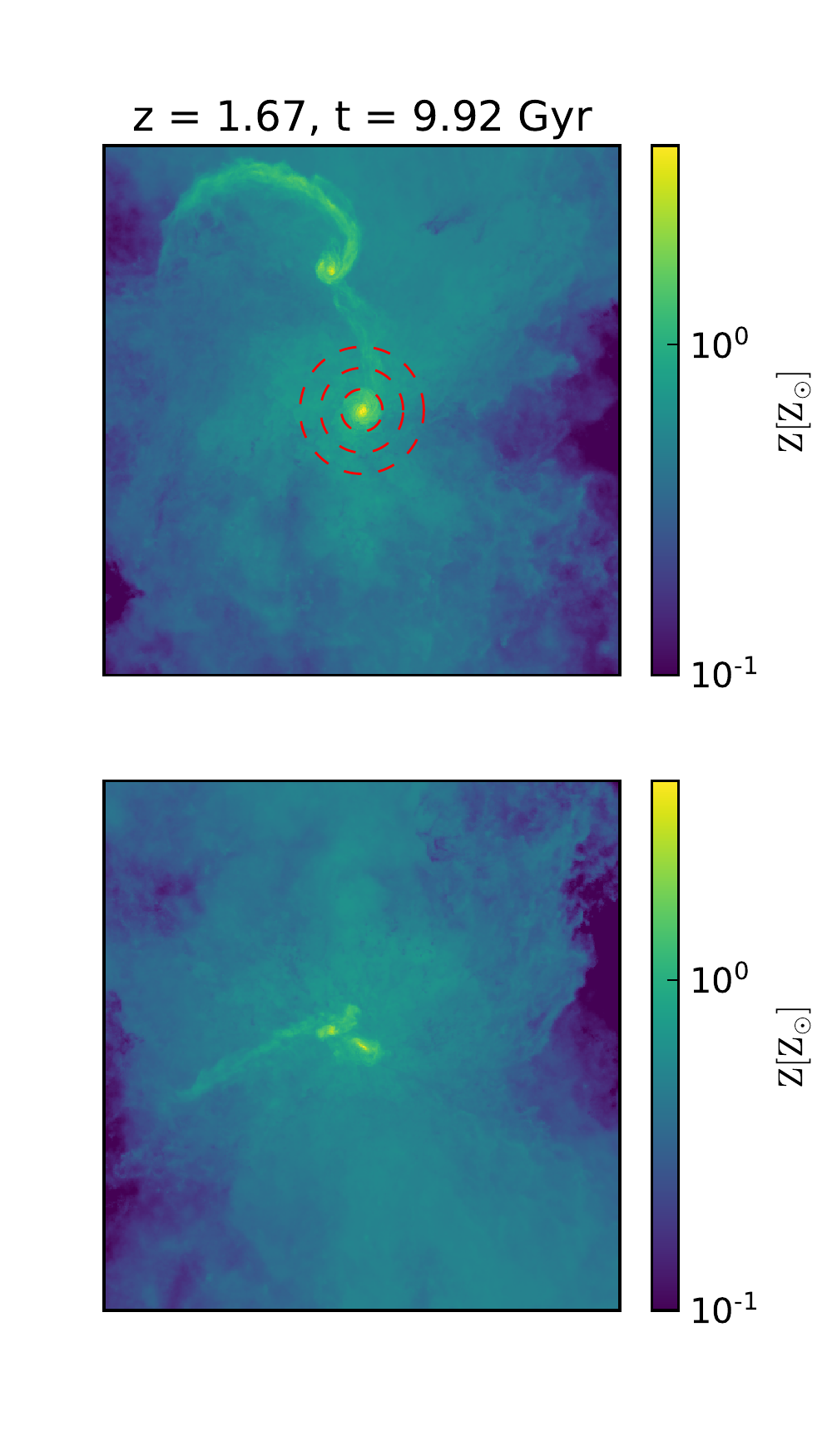}
\includegraphics[scale=0.6,trim={1.cm 1.cm 2.5cm 1.75cm},clip]{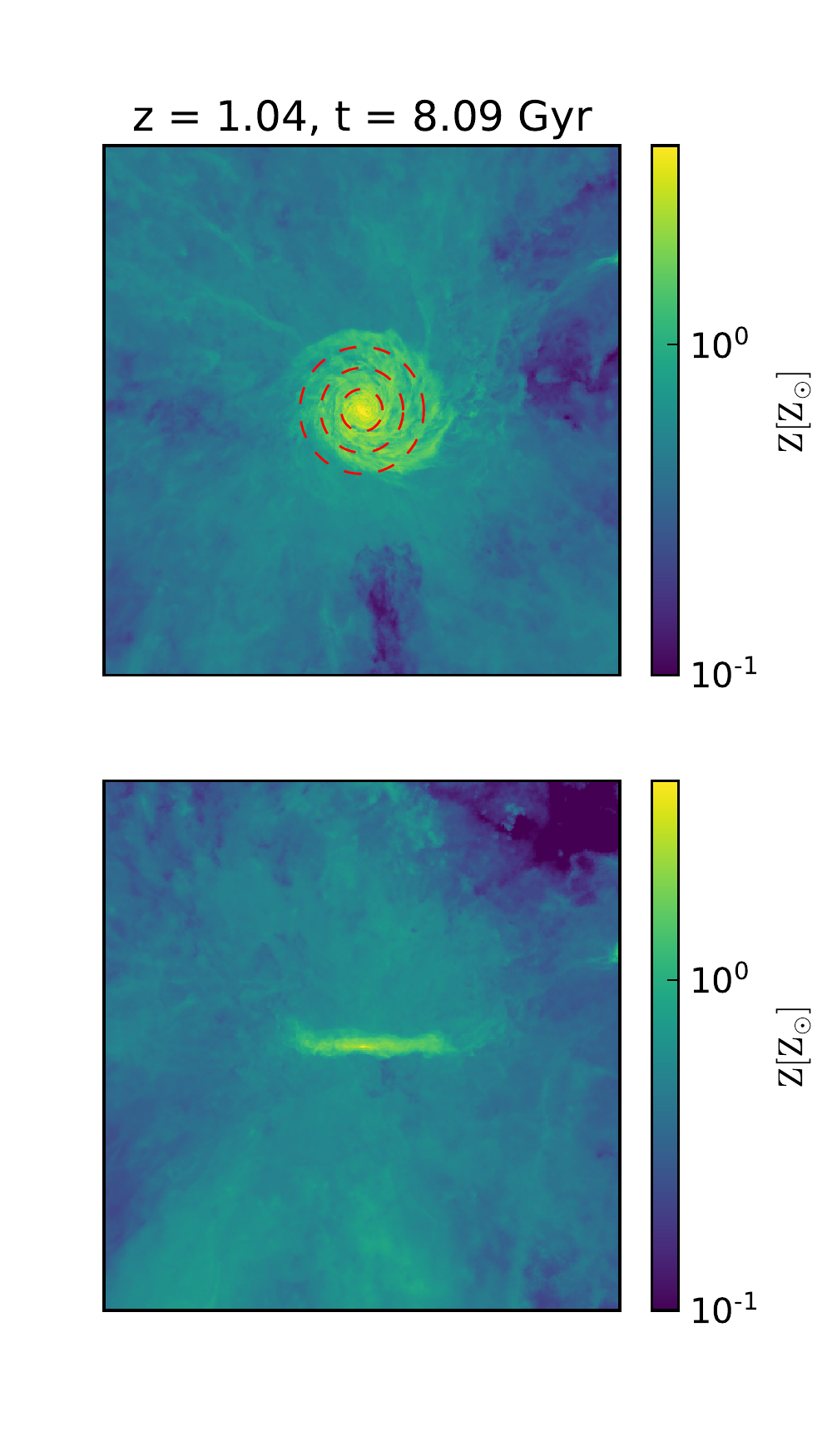}
\includegraphics[scale=0.6,trim={1.cm 1.cm 2.5cm 1.75cm},clip]{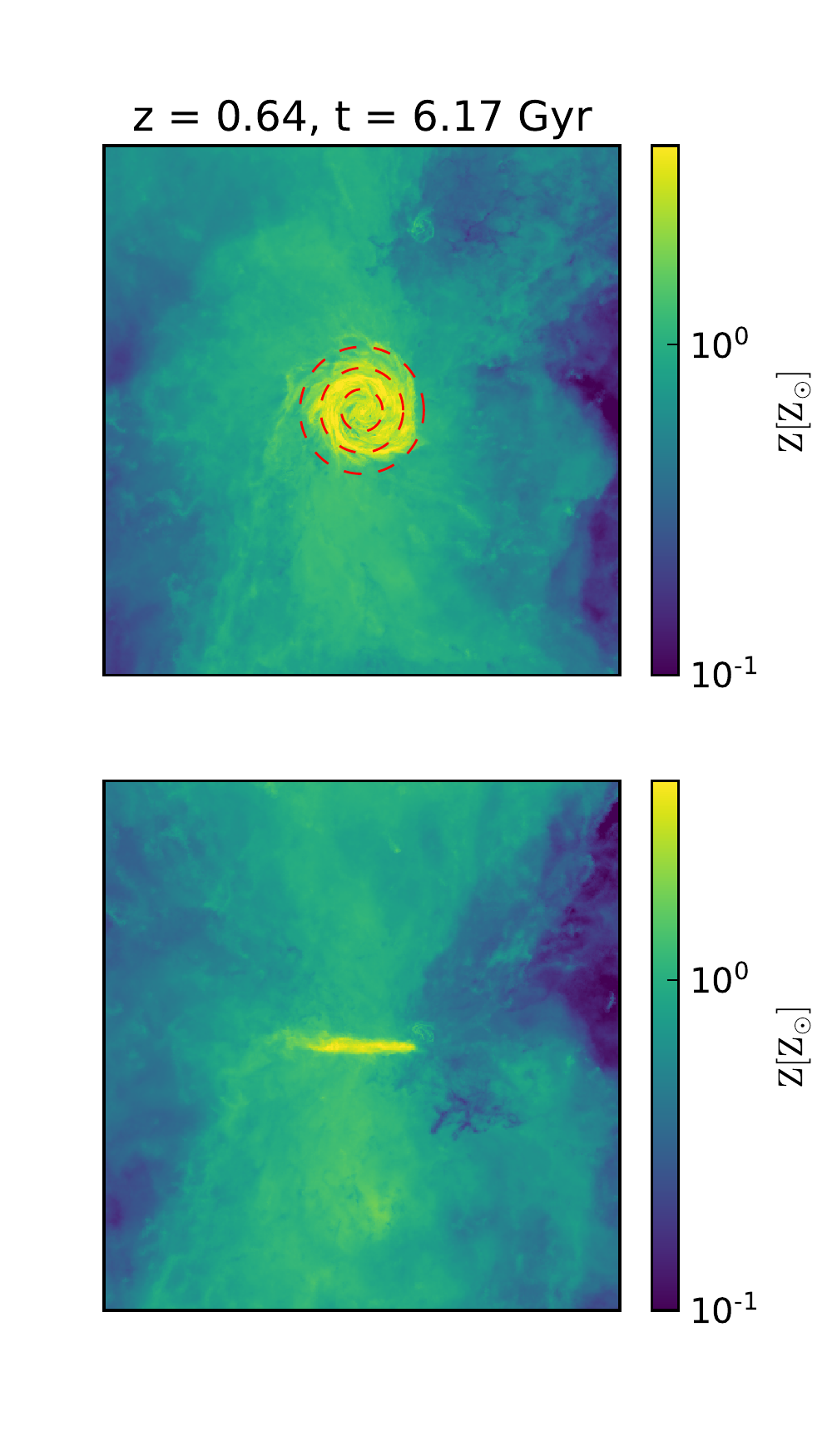}
\includegraphics[scale=0.6,trim={1.cm 1.cm 0.2cm 1.75cm},clip]{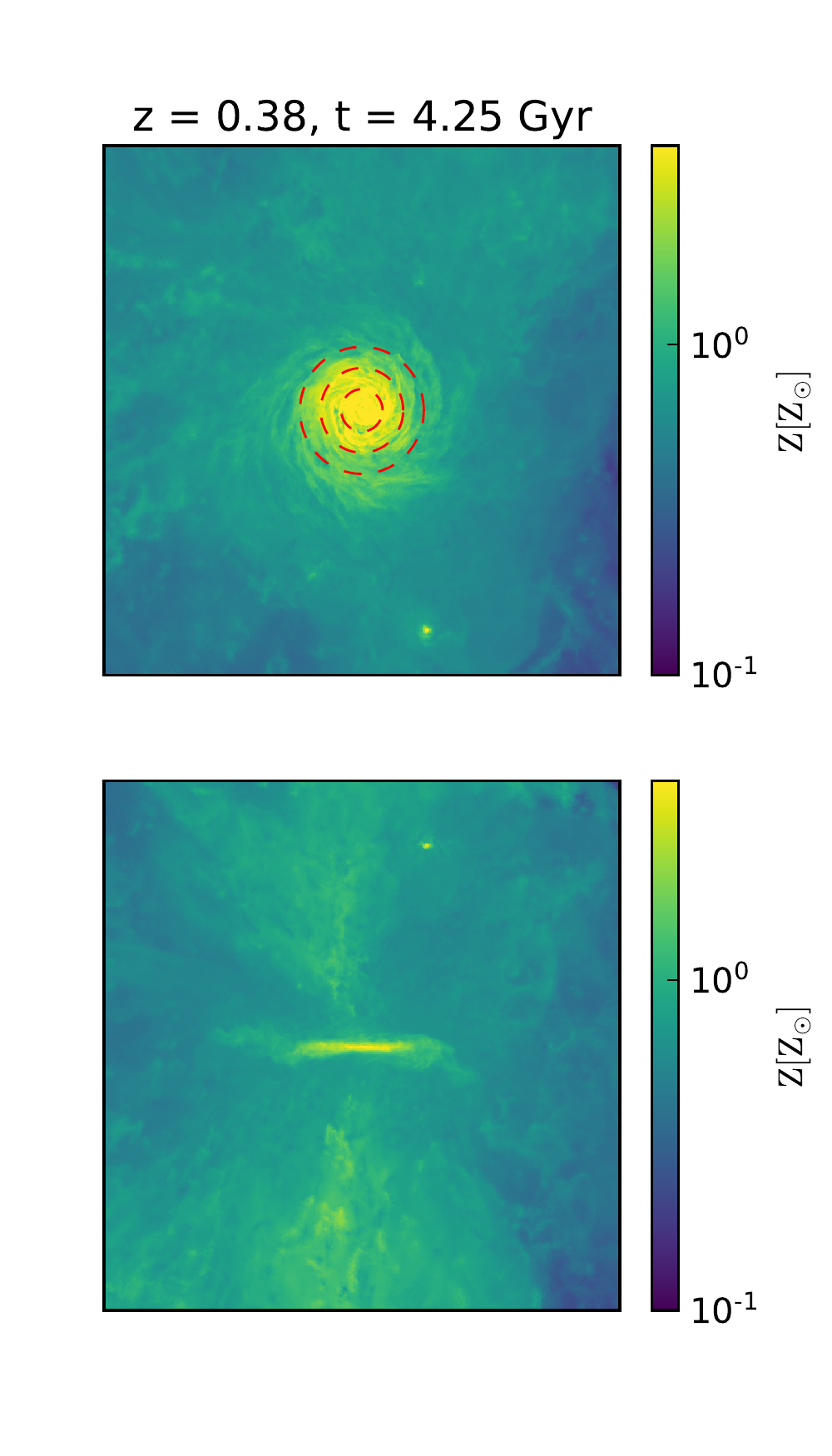}\\
\caption{Projections of the face-on and edge-on gas density distribution (\emph{top panels}) and the gas metallicity distribution (\emph{bottom panels}) for simulation Au 23. Dashed circles mark radii of 4 kpc, 8 kpc and 12 kpc. The snapshots are spaced $\sim 2$ Gyr apart, and show the gas-rich major merger period (\emph{1st column}); the end of the high-$\alpha$ sequence formation at $R=12$ kpc (\emph{2nd column}); the shrunk gas disc (\emph{3rd column}); and the subsequent inside-out growth of the second generation of low-$\alpha$ stars (\emph{4th column}). }
\label{fig3}
\end{figure*}

\begin{figure*}
\includegraphics[scale=1.5,trim={1.cm 0 .1cm 0},clip]{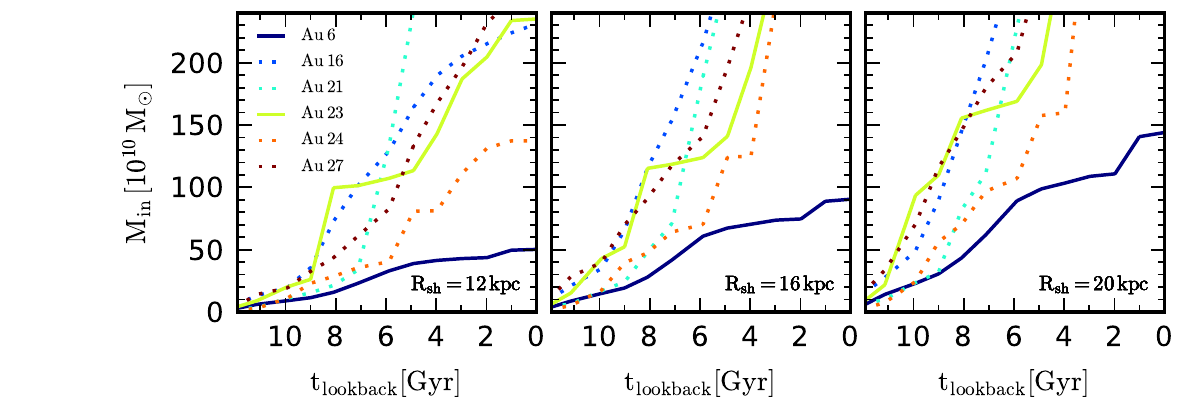}
\caption{The cumulative gas mass that has flowed inward through three spherical shells, for all simulations. Solid lines indicate haloes in which the gas disc undergoes a transient period of significant shrinkage. Lines are coloured according to the labels in Fig.~\ref{fig2}.}
\label{gacc}
\end{figure*}

Following \citet{BRS15}, we define $\alpha = \rm (O+Mg+Si) / 3$\footnote{We do not follow the evolution of Sulphur and Calcium, therefore we do not include these elements in the average $\alpha$ definition.}. In Fig.~\ref{fig1}, we show the distribution of star particles in the $\rm [\alpha/Fe]$--$\rm [Fe/H]$ plane at different radii, overlaid with contours of mean age, for all six simulations. In each case, we observe the expected result that older stars have higher $\rm [\alpha/Fe]$ and lower $\rm[Fe/H]$ than younger star particles. We note that the distribution in chemical abundance space varies between radii and simulations; the distributions are either smooth or comprise two distinct sequences in $\rm [\alpha/Fe]$, i.e., a chemical dichotomy, consisting of a high- and low-$\rm [\alpha/Fe]$ sequence located at around $\rm [\alpha/Fe]\sim0.1$ and $\sim0.02$, respectively. In the inner regions ($R<5$ kpc), simulations Au 16, Au 23, Au 24 and Au 27 show a clear dichotomy, whereas the two sequences in Au 6 and Au 21 are connected to each other, resulting in a less clear dichotomy. In the outer regions ($R\gtrsim7$ kpc), only simulation Au 23 shows a dichotomy with a clear gap between the high- and low-$\rm [\alpha/Fe]$ sequences. In the other simulations, the high-$\rm [\alpha/Fe]$ disc does not extend farther than $R\sim 7$ kpc, therefore the outer disc houses only the low-$\rm [\alpha/Fe]$ sequence. In general, the high-$\rm [\alpha/Fe]$ sequence is more radially compact than the low-$\rm [\alpha/Fe]$ disc, in agreement with what is observed in the Milky Way \citep[e.g.][]{BAO11,HBH15}. We verified that these distributions are unchanged for $\emph{in-situ}$ stars only, therefore $\emph{ex-situ}$ material that may be accreted into a disc configuration \citep{GGM17b} is not a dominant mechanism for chemical dichotomy formation.

\subsection{Pathways to a chemical dichotomy}

In this section, we show that all the simulations have a strong high-$\rm[\alpha/Fe]$ star formation phase in the inner region followed by a more quiescent low-$\rm[\alpha/Fe]$ disc growth phase. Our galaxies show a variety of intensity, duration and size of the first star formation phase that together determine the strength of the dichotomy in the inner region. The simulated galaxies also show different histories of the low-$\rm[\alpha/Fe]$ disc growth. The radial extent of the high-$\rm[\alpha/Fe]$ disc relative to that of the initial low-$\rm[\alpha/Fe]$ disc determines the strength of the dichotomy in the outer region ($R\gtrsim7$ kpc), which indicates a dependence on the disc growth history. Hence, there are two pathways in which a clear dichotomy can be created: one which applies in the inner region ($R\lesssim 5$ kpc), and the other applies to the outer region ($R\gtrsim 7$ kpc).

\subsubsection{Centralised starburst pathway: inner disc dichotomy}

We show in the top-row of Fig.~\ref{fig2} the star formation histories (SFHs) in four radial regions\footnote{Note that the SFH is measured from the star formation rate of all gas cells within the given radial range at a given time, and not inferred from stellar ages at present day.}, for all simulations, using time-bins of 1 Gyr width. We note that simulations Au 16, Au 23, Au 24 and Au 27 have strong bursts of star formation in their inner regions ($R<3$ kpc in Au 23 and Au 24 and $R<5$ kpc in Au 16 and Au 27) at $t_{\rm lookback}\sim 10$ Gyr. This star formation lasts for approximately 2 Gyr, after which time the SFR drops to a low level. The middle- and bottom-rows of Fig.~\ref{fig2} show the evolution of the median and root mean square of the gas $\rm [\alpha/Fe]$ and $\rm [Fe/H]$ distributions, respectively, at the radii corresponding to those of the SFHs. In addition, the lower panels of Fig.~\ref{fig2} indicate the times of gas-rich mergers (listed in Table. 1) and flybys for each simulation.

We note that the short periods of intense star formation in these simulations are caused by gas-rich major mergers at $t_{\rm lookback}\gtrsim8$ Gyr, which produce stars from gas that maintains a roughly constant value of $\rm [\alpha/Fe]\sim 0.1$ dex that populate the high-$\rm[\alpha/Fe]$ sequence. As SNeIa enrichment becomes important $\sim 1$ Gyr later, the gas rapidly transitions to lower values of $\rm [\alpha/Fe]\sim 0.02$ dex and higher $\rm[Fe/H]$ as the burst ends, which becomes star-forming material for the low-$\rm[\alpha/Fe]$ sequence that builds up during the subsequent period of lower SFR. Owing to this rapid transition and relatively low level of star formation, the transition region is populated with relatively few stars, therefore the subsequent star formation that follows from low-$\rm[\alpha/Fe]$ gas after the transition provides the over-density at lower (higher) $\rm[\alpha/Fe]$ ($\rm [Fe/H]$)\footnote{This is naturally expected in one zone models and the same scenario is demonstrated in the inner region of the chemical evolution model of \citet[][]{SB09b}. Note that in this process, the low-$\rm[\alpha/Fe]$ sequence starts from left-over gas of the high-$\rm[\alpha/Fe]$ sequence. Hence, unless there is massive inflow of low-$\rm[Fe/H]$ gas (which does not occur in the inner regions in our simulations), the low-$\rm[\alpha/Fe]$ sequence has higher $\rm[Fe/H]$ than the high-$\rm[\alpha/Fe]$ sequence.}. This scenario is clearest in the inner regions for Au 16 and Au 23, in which the gas remains at $\rm [\alpha/Fe]\sim 0.02$ dex for about 2 Gyr (after the concentration of high-$\rm[\alpha/Fe]$ stars formed from the initial rapid star formation), during which time star formation builds a concentration of low-$\rm[\alpha/Fe]$ stars. We note that in simulations Au 6, Au 21 and Au 27, the SFH in the inner regions is weaker at early times - which produces fewer high-$\rm[\alpha/Fe]$ stars, and more prolonged than the intense burst in Au 16 and Au 23. This leads to a significant amount of stars formed from gas that moves from high-$\rm[\alpha/Fe]$ to low-$\rm[\alpha/Fe]$ and high-$\rm[Fe/H]$. As a result, the transition region between the two sequences is populated such that the two sequences join at the same $\rm[Fe/H]$ value, and no clear dichotomy can be discerned.

\subsubsection{Shrinking disc pathway: outer disc dichotomy}

In the outer region ($R\gtrsim7$ kpc), only Au 23 shows a clear dichotomy in Fig.~\ref{fig1}. This dichotomy was created from a different pathway to that of the inner region. The SFH at $R=12$ kpc reveals two peaks of SF: the first peak occurs at $t_{\rm lookback}\sim 8$ Gyr; the second at $t_{\rm lookback}\sim 4$ Gyr. Between these peaks the SFR drops nearly to zero, during which time the $\rm[\alpha/Fe]$ of the gas rapidly decreases from $\sim 0.08$ dex to $\sim 0.02$ dex. The dearth in the SFH at $t_{\rm lookback}\sim 6$ Gyr is caused by a temporary reduction in the size of the gas disc immediately after it grew to $R\sim12$ kpc at  $t_{\rm lookback}\sim 8$ Gyr, which created the high-$\rm[\alpha/Fe]$ disc stars. Fig.~\ref{sfg} shows the evolution of the gas disc `edge', defined to be the radius at which the gas surface density falls below $5$ $\rm M_{\odot}$ $\rm pc^{-2}$. We note that this value is approximately the threshold value beyond which the gas in our simulations is star-forming: the star formation model of \mbox{\citet{SH03}} is calibrated to the relation given by \citet{K89} such that star formation rates are very low for surface densities below 10 $\rm M_{\odot}$ $\rm pc^{-2}$. Furthermore, the HI gas surface density radial profiles of our simulations \citep[presented in][]{MGP16} drop steeply after reaching a surface density of around $5$ $\rm M_{\odot}$ $\rm pc^{-2}$, which can be considered the edge of the star-forming gas disc. It is clear from Fig.~\ref{sfg} that the gas disc grew from $R\sim 5$ kpc at $t_{\rm lookback}\sim 10$ Gyr to $R\sim 15$ kpc at $t_{\rm lookback}\sim 8$ Gyr, then decreased in size to $R\sim 10$ kpc at $t_{\rm lookback}\sim 6$ Gyr, before resuming growth and reaching $R\sim 15$ kpc again at $t_{\rm lookback}\sim 4$ Gyr. This evolution is also visualised in the top panels of Fig.~\ref{fig3}. 

The shrinking\footnote{We stress that the term `shrinking' refers to the decrease in size of the gas disc only, and does not imply the bulk inward motion of gas.} of the gas disc allows a halt in SF at $R=12$ kpc while SNeIa enrich the gas at this radius to lower $\rm[\alpha/Fe]$ before SF continues when the gas disc grows again to this radius. In this particular case, the gas $\rm [Fe/H]$ decreases before the second peak in SF, which allows the low-$\rm[\alpha/Fe]$ sequence to begin at lower metallicity than the end of the high-$\rm[\alpha/Fe]$ sequence, thus forming a bimodal distribution in $\rm[\alpha/Fe]$ at $\rm [Fe/H]\sim 0.0$ dex. This is in stark contrast to the dichotomy pathway in the inner region, in which the low-$\rm[\alpha/Fe]$ sequence always starts at higher $\rm[Fe/H]$ than the high-$\rm[\alpha/Fe]$ sequence, i.e. there is no bimodality in $\rm[\alpha/Fe]$ for any $\rm[Fe/H]$. A similar SFH is observed for Au 6 at $R=8$ kpc. However, as mentioned above, in Au 6 the initial star formation phase is prolonged, and the high-$\rm[\alpha/Fe]$ sequence is merged to the low-$\rm[\alpha/Fe]$ sequence. Hence, the dichotomy is far less clear. 

Fig.~\ref{sfg} shows that the other simulations exhibit a more continuous inside-out growth of their discs, which is also seen in the SFHs that move progressively outward with time in the top-row of Fig.~\ref{fig2}. This has two implications: i) the early star formation phase is concentrated only in the inner regions ($R\lesssim 5$ kpc), which does not produce a population of high-$\rm[\alpha/Fe]$ in the outer region\footnote{Unless radial migration brings them from the central region to the outer region. However, radial migration does not appear to play an important role in forming the dichotomy, which we discuss in Section. 4.}. As a result, simulations Au 21, Au 24 and Au 27 show only the low-$\rm[\alpha/Fe]$ sequence in the outer disc. ii) There is no disc shrinking at any time. This means that even if the high-$\rm[\alpha/Fe]$ sequence extends to $R\sim10$ kpc, as in Au 16 (see Fig.~\ref{fig1}), there is not a significant decrease in star formation or lowering of the gas metallicity to create a clear dichotomy extending to a lower $\rm[Fe/H]$, as in Au 23. Therefore, to have a chemical dichotomy in the outer region, the high-$\rm[\alpha/Fe]$ sequence must be built quickly ($\lesssim 3$ Gyr) from the central regions up to $R\gtrsim 7$ kpc during a period of early, intense star formation. This is then followed by a clear reduction of star formation rate (caused by a shrinking of the gas disc) that drives a quick decrease in $\rm[\alpha/Fe]$, with subsequent gas accretion that forms the low-$\rm[\alpha/Fe]$ disc in an inside-out fashion.

\begin{figure*}
\includegraphics[scale=1.1,trim={.5cm 0.5cm 0cm 0},clip]{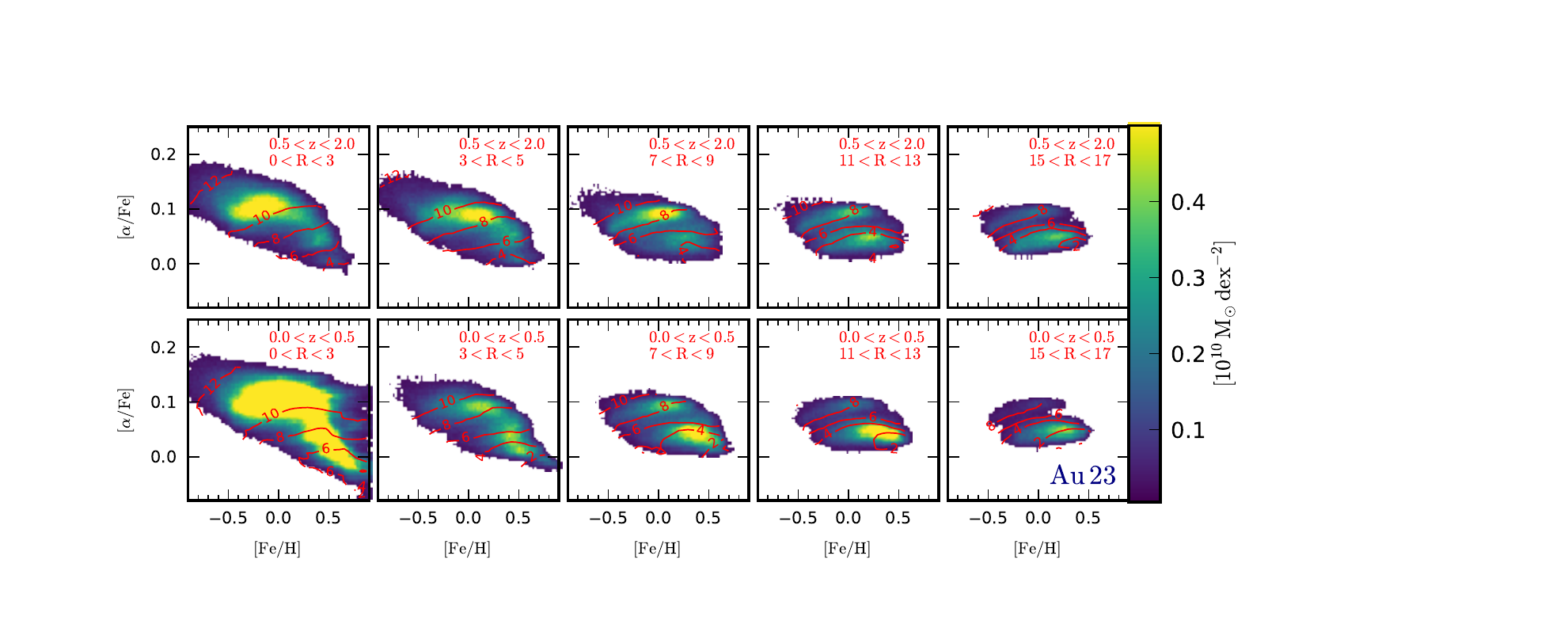}\\
\includegraphics[scale=1.1,trim={.5cm 0.5cm 0cm 1.5cm},clip]{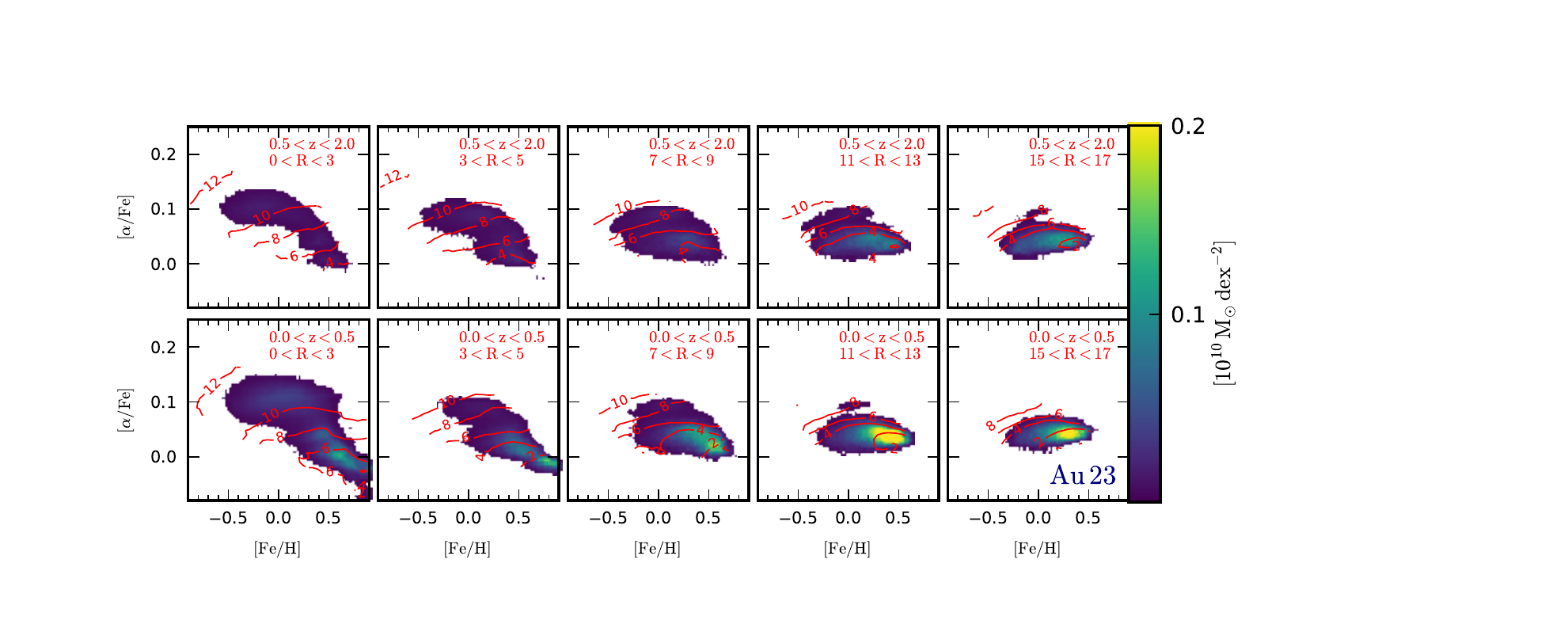}\\
\caption{\emph{Top}: As Fig.~\ref{fig1}, but for radii at different heights from the midplane for simulation Au 23. \emph{Bottom}: As the top panels, but with the APOGEE red clump star age-weighted selection function \citep[see equation (11) of][]{BNW16}. Each panel follows the scale marked on the colour bar.}
\label{au23}
\end{figure*}

\begin{figure*}
\includegraphics[scale=1.1,trim={0cm 1.5cm 0cm 0},clip]{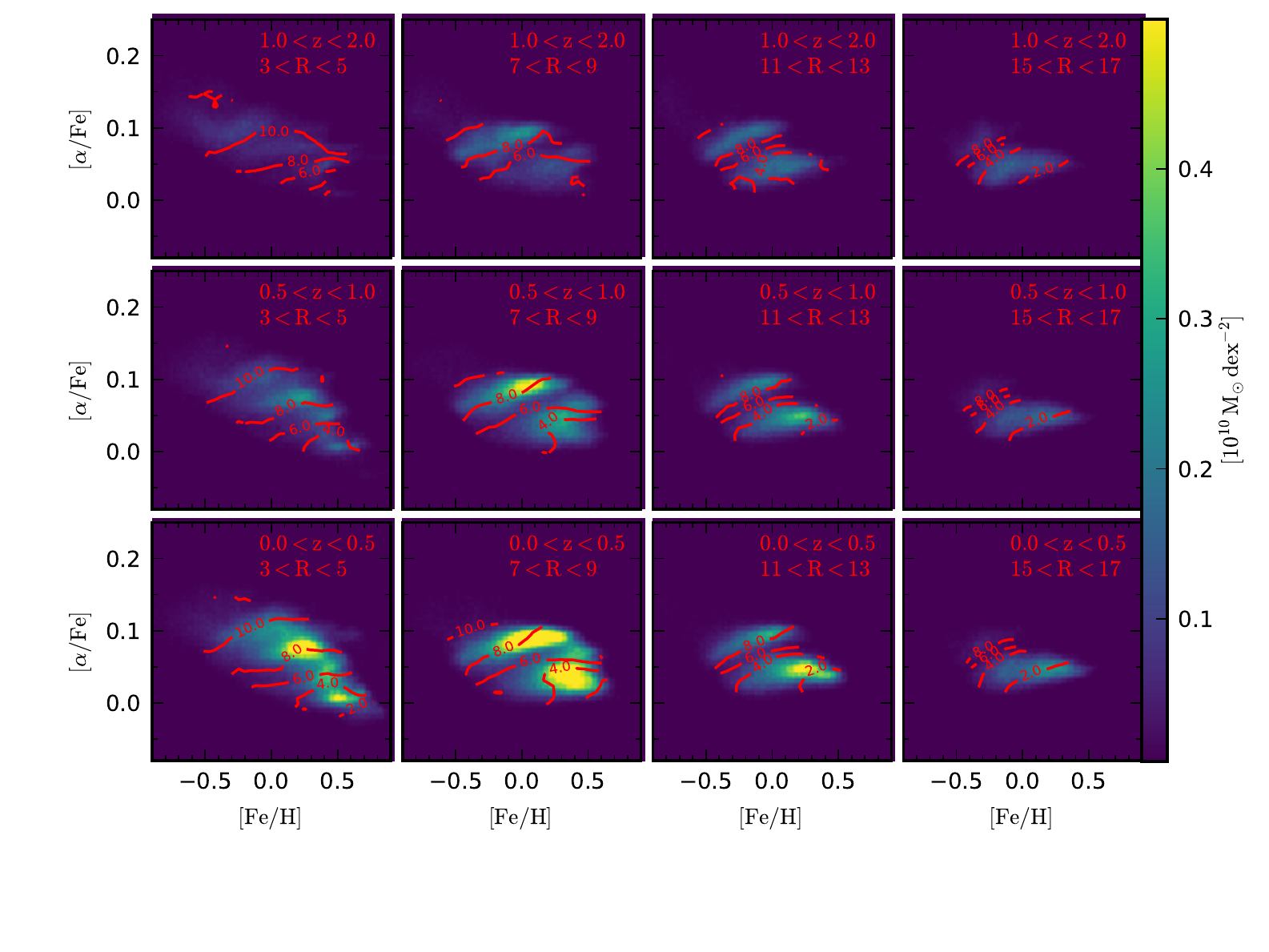}
\caption{As Fig.~\ref{au23}, but for the star particles at their birth positions.}
\label{au23b}
\end{figure*}

\subsection{Gas disc evolution}

As discussed above, the formation of the chemical dichotomy in the outer disc region appears to rely on the temporary shrinking of the gas disc while the gas evolves to a low-$\rm[\alpha/Fe]$ state. To study the reasons behind the gas disc shrinking, we show in Fig.~\ref{gacc} the cumulative gas inflow mass for three spherical shells of 4 kpc width, centred on radii of 12 kpc, 16 kpc and 20 kpc, for all simulations. For the simulations in which the gas disc shrinks (Au 6 and Au 23), significantly low gas inflow rates are seen at all of these radii at $t_{\rm lookback}\sim6$-9 Gyr, whereas the gas inflow rates either increase or maintain a steady value for the other simulations. This indicates that, for Au 6 and Au 23, the gas accretion at these times is not sufficient to replenish the gas consumed by star formation around the disc edges, which leads to the shrinking of the gas disc until $t_{\rm lookback}\sim6$ Gyr when gas accretion becomes significant enough to grow the disc. We note that the stellar disc does not shrink with the gas disc, which indicates that tidal forces from satellites are not a driver of gas disc shrinking.

At late times, the hot gas phase is already in place and the halo does not allow gas filaments to penetrate into the central regions \citep{KKW05,DB06,BGQ09,FJP12}. An important source of cold gas accretion at this late epoch appears to be minor gas-rich mergers that are able to penetrate the halo and reach the central galaxies. On their way to the centre, their gas is stripped and coalesces around the edge of the disc, allowing it to grow. This is shown in Fig.~\ref{fig3} for Au 23, in which the gas disc does not grow again until a satellite galaxy penetrates into the inner galaxy. Moreover, the gas accreted from the minor merger is more metal poor than that of the central galaxy (see lower panels of Fig.~\ref{fig3} and Table.~1), which leads to the temporary dip in gas metallicity around the disc edge as discussed in the previous sub-section and shown in the lower panels of Fig.~\ref{fig2} (see also Bustamante et al. in prep). This helps to create low-$\rm[\alpha/Fe]$ stars at lower $\rm[Fe/H]$ than the highest $\rm[Fe/H]$ stars of the high-$\rm[\alpha/Fe]$ sequence. On the other hand, the lower panels of Fig.~\ref{fig2} show that haloes Au 16 and Au 21, which undergo continuous inside-out growth, experience regular interactions with gas-rich satellites (we have verified this from the simulation snapshots). The galaxies in these simulations therefore have a constant supply of cold gas out of which the discs grow at all times. We note that generally, but especially in the simulations with smooth inside-out growth, the gas metallicity continues to increase with time even at late times. This is responsible for the banana shape in the low-$\rm[\alpha/Fe]$ sequence. In the Milky Way, the age-metallicity relation is expected to be flat, therefore the low-$\rm[\alpha/Fe]$ sequence is likely kept at similar $\rm[Fe/H]$ at a given radius \citep[see Fig. 4 of][]{HBH15}. The reason for the increasing gas metallicity in the simulations is unclear, and we leave this issue and its implications to a detailed future study.

\section{Conclusions and Discussion}
\label{sec4}

In this study, we have used a set of high resolution cosmological zoom simulations to analyse the evolution of the stars at different spatial regions in chemical space, with particular focus on the formation of two distinct sequences in $\rm[\alpha/Fe]$--$\rm[Fe/H]$, a chemical-dichotomy. Our main results are as follows:

\begin{itemize}
\item{} There are two pathways to forming a clear chemical dichotomy: an early \emph{centralised starburst} mechanism that is relevant for the inner disc; and a \emph{shrinking disc} mechanism that is relevant for the outer disc. 
\item{} For the \emph{centralised starburst} pathway: an early, gas-rich merger induced starburst creates stars at high-$[\rm\alpha/Fe]$ and high $\rm[Fe/H]$, which is followed by a low level of star formation that forms the low-$\rm[\alpha/Fe]$ sequence. The rapid transition from high- to low-$\rm [\alpha/Fe]$ ensures a depopulated region between the two sequences. 
\item{}For the \emph{shrinking disc} pathway: at later times ($\rm t_{lookback}\lesssim 8$ Gyr), the gas disc shrinks after the high-$\rm[\alpha/Fe]$ sequence forms, causing a decrease in SFR while the gas transitions to a low-$\rm[\alpha/Fe]$ state. Subsequent low-metallicity gas accretion grows the low-$\rm[\alpha/Fe]$ sequence in an inside-out fashion from lower $\rm[Fe/H]$ than the end of the high-$\rm[\alpha/Fe]$ sequence. 
\item{} The main source of gas at $\rm t_{lookback}\lesssim 8$ Gyr is from gas-rich, minor mergers that penetrate the hot-mode dominated halo and deliver stripped gas to the main galaxy. 
\item{} In general, the gas phase metallicity continuously increases with time at a given radius, even at late times.
\end{itemize}

Our study, as well as many previous studies of numerical simulations, is consistent with what is suggested in \citet{BKG04}, who showed that a violent, gas-rich merger can produce a thick and compact disc in the earlier epoch, followed by a more quiescent gas accretion and inside-out disc growth \citep{BGM05,BKM06}. However, none of the previous works clearly explained the origin of the chemical dichotomy seen in their simulations. Owing to our very high-resolution cosmological simulations, we have finally identified the origin of this dichotomy, and have clearly demonstrated that there are two different pathways to the dichotomy that operate in different regions of the disc.

We note that in the shrinking disc pathway, the high-$\rm[\alpha/Fe]$ sequence forms inside-out \citep[see also][]{KAB17} up to radii of $\sim 12$ kpc within $\sim 3$ Gyr. This is contrary to the notion that the chemical thick disc formed with a spatially homogeneous star formation history put forward by some chemical evolution models \citep[][]{SHD15}. With regard to two-infall models, however, the temporary drop in gas accretion is remarkably reminiscent of the two distinct periods of gas infall discussed in \citet{CMG97} and \citet{CM01}. However, the secondary gas accretion in the simulation is not primordial, and the timescale of the chemical thick disc formation is not as short as found in that work (although the timescales are comparably short given the lack of thick disc metallicity constraints 20 years ago).

The inside-out formation of the low-$\alpha$ sequence that occurs after the shrinking of the large high-$\rm[\alpha/Fe]$ star-forming gas disc appears to account for the clear dichotomy in the region $7\lesssim R\lesssim13$ kpc, at which the two sequences significantly overlap in $\rm[Fe/H]$ (a bimodality in $\rm[\alpha/Fe]$). In addition, the outer regions are heavily dominated by the low-$\rm[\alpha/Fe]$ sequence. This is shown explicitly in Fig.~\ref{au23} for the clearest case of Au 23, in which we show the chemical abundance plane at a series of radii \emph{and heights} with a common normalisation, for all star particles (top panels) and for the APOGEE Red Clump age-dependent selection function \citep[see equation (11) of][]{BNW16} (bottom panels). For all star particles, we see that for $3<R<5$ kpc in the plane, the distinct low- and high-$\rm[\alpha/Fe]$ sequences exist in equal parts, with the former increasing in dominance over the latter with increasing radii. The low-$\rm[\alpha/Fe]$ sequence fades with increasing distance above the plane in the inner radii, but becomes more prominent at larger radii to dominate over the high-$\rm[\alpha/Fe]$ sequence which fades at larger radii for all heights. It is interesting to note that the application of the age-dependent selection function (bottom panels of Fig.~\ref{au23}), the high-$\rm[\alpha/Fe]$ population is much less dominant than in the top panels of Fig.~\ref{au23}: it is visible only in the innermost regions, and almost not at all visible beyond $R\sim 9$ kpc. These trends are consistent with two distinct chemical components that qualitatively reproduce the vertically thick, radially compact high-$\rm[\alpha/Fe]$ disc and the vertical thin, flaring and radially extended low-$\rm[\alpha/Fe]$ disc as claimed to be seen in observations \citep[e.g.][]{BAO11,RCK13,MMN16}. We note also that the low-$\rm[\alpha/Fe]$ sequence follows a negative radial metallicity gradient (which is present in the gas at all times, see Fig.~\ref{fig2}), and is consistent with the `donut' distribution of MAPs noted in \citet{BRS15}.

Comparison between Fig.~\ref{au23} and what is observed in the Milky Way \citep[e.g.][]{NBB14,HBH15} is interesting\footnote{We note that \citet{HBH15} studied the distribution of APOGEE giant stars - not only the Red Clump sample. Therefore, our comparison with the results of this study should be regarded in a qualitative sense only.}. The APOGEE data shows that, in the inner disc ($R<5$ kpc), the low-$\rm[\alpha/Fe]$ sequence is not extended to low $\rm[Fe/H]$, but instead the low-$\rm[\alpha/Fe]$ sequence is more metal rich than (or an extension of) the high-$\rm[\alpha/Fe]$ sequence, rather similar to our \emph{centralised starburst} pathway. Around the solar radius of $R\sim8$ kpc, however, the two sequences overlap in $\rm[Fe/H]$: the low-$\rm[\alpha/Fe]$ sequence extends to lower $\rm[Fe/H]$ and makes a clear gap in $\rm[\alpha/Fe]$. This is similar to what our \emph{shrinking disc} pathway produced in Au 23 in the outer region. Interestingly, \citet{RRF14} suggest the shrinking of a younger thick disc from their stellar population analysis of SDSS and 2MASS data, which may indicate that their young thick disc corresponds to our shrunk initial low-$\rm[\alpha/Fe]$ disc. Hence, we expect that both of these pathways occurred in the Milky Way.

It is interesting to note that the features in chemical abundance space for the star particles are in place at birth (see Fig.~\ref{au23b}), which implies that this formation scenario is consistent with a flared star-forming gas disc (increasing scale height with radius) that is argued to be required to produce the observed negative vertical metallicity gradients of coeval populations \citep{XLY15,KGG17,CKL17}. Another implication is that processes such as radial migration are not required in order to \emph{form} the chemical dichotomy if the high-$\rm[\alpha/Fe]$ disc is as large as in Au 23, but rather influence the detailed structure of the features through radial mixing \citep{GKC15,LDN16,GSK16} and heating \citep{HBH15}. These effects are beyond the scope of this paper.

Further implications for the Milky Way are the importance of the gas accretion history at late times, which seems to be dominated by gas-rich minor mergers in these simulations. This may indicate that, if the dichotomy observed in the Milky Way was formed by processes similar to the shrinking gas disc scenario discussed in this study, the merger history of the Galaxy ought to have been quiet during the period just after the early gas rich merger and when the hot gas halo has grown enough such that hot-mode accretion becomes dominant. This would allow the gas inflow to slow-down before the disc grew to its present size. 

We stress that Au 23 and the other simulations in this study are not tailored to exactly match the properties of the Milky Way, which is reflected in chemical structure of the simulated discs: the precise loci and shapes of the two sequences in chemical space are different from that in the Milky Way. We note also that the mean stellar iso-age contours in the $\rm[\alpha/Fe]$--$\rm[Fe/H]$ plane tend to be quite flat, i.e., they do not have a negative slope as observed in APOGEE data \citep[see][]{NHR16}. The shape of these features are probably related to the interplay between SNII and SNIa, controlled mainly by the delay time distribution and normalisation of SNeIa in our simulations \citep[explored in the Aquarius simulations in][]{MPS14b}, but also by the initial mass function \citep[see][for a study of the effects of variable IMFs on the {$\rm[\alpha/Fe]$--$\rm[Fe/H]$} plane]{GS17}. A detailed analysis of the stellar distribution in abundance and age space has the potential to place tight quantitative constraints on the assembly history of the Milky Way, and will be the subject of future studies.

The enrichment history seen in the simulations contrasts with the expectations of many analytic models: the simulations show that the gas metallicity continuously increases with time, whereas analytic models expect that, in the Milky Way, the gas metallicity reaches a constant value at all radii on timescales that increase with radius. This may indicate that the expected equilibrium between metal enrichment from stellar feedback and dilution of the ISM from gas infall is not reached in the simulations. If this is not the case for the Milky Way, this will cause differences between our simulations and observations in the detailed chemical structure and in particular the locus of the low-$\rm[\alpha/Fe]$ sequence in chemical abundance space. We will investigate the cause of this result in future studies.

Additional dependencies include the merger history, the rate of inside-out formation and the amount of radial migration, the last of which has further dependencies on the bar and spiral structure of the disc \citep[e.g.][]{GKC11,GKC12,MFS16b} and therefore the dynamical heating history, which includes also external effects. Given this complexity, and that the simulations presented in this study do not match the Milky Way in every aspect, we do not expect the simulations to match the detailed structure of the Milky Way, but rather provide a cosmologically motivated insight into the mechanisms that may shape certain features of our Galaxy, in this case, its chemical structure. In the future, detailed analysis of simulations with more constrained initial conditions aimed at reproducing the Local Group environment will bring cosmological simulations more inline with the Galaxy, and provide important clues to its formation.

\section*{acknowledgements}
We thank the referee, as well as Cristina Chiappini and Ivan Minchev, for useful comments that helped improve the manuscript. RG and DK acknowledge discussions held at the IAU symposium 334 `Rediscovering our Galaxy', which helped encourage this study. RG and VS acknowledge support by the DFG Research Centre SFB-881 `The Milky Way System' through project A1. DK acknowledges the support of the UK's Science \& Technology Facilities Council (STFC Grants ST/N000811/1). This work has also been supported by the European Research Council under ERC-StG grant EXAGAL- 308037. Part of the simulations of this paper used the SuperMUC system at the Leibniz Computing Centre, Garching, under the project PR85JE of the Gauss Centre for Supercomputing. This work used the DiRAC Data Centric system at Durham University, operated by the Institute for Computational Cosmology on behalf of the STFC DiRAC HPC Facility `www.dirac.ac.uk'. This equipment was funded by BIS National E-infrastructure capital grant ST/K00042X/1, STFC capital grant ST/H008519/1, and STFC DiRAC Operations grant ST/K003267/1 and Durham University. DiRAC is part of the National E-Infrastructure.

\bibliographystyle{mnras}
\bibliography{GG4d5R1.bbl}

\label{lastpage}

\end{document}